\documentclass[aps,floatfix,showpacs,nofootinbib,twocolumn]{revtex4-1}
\usepackage{amsmath}
\usepackage{amsfonts}
\usepackage{amssymb}
\usepackage{graphics,epsfig}
\usepackage{graphicx}
\begin{document}
\title{The McVittie solution with a negative cosmological constant}
\author{Philippe Landry}
\email[]{plandry@uoguelph.ca}
\author{Majd Abdelqader }
\email[]{majd@astro.queensu.ca}
\author{Kayll Lake}
\email[]{lake@astro.queensu.ca}
\affiliation{Department of Physics, Queen's University, Kingston,
Ontario, Canada, K7L 3N6 }
\date{\today}

\begin{abstract}
Whereas current cosmological observations suggest that the universe is dominated by a positive cosmological constant ($\Lambda > 0$), the AdS/CFT correspondence tells us that the case $\Lambda<0$ is still worthy of consideration. In this paper we study the McVittie solution with $\Lambda<0$. Following a related study, the solution is understood here by way of a systematic construction of conformal diagrams based on detailed numerical integrations of the null geodesic equations. As in the pure Robertson - Walker case, we find that $\Lambda<0$ ensures collapse to a Big Crunch, a feature which completely dominates the global structure.
\end{abstract}
\maketitle

\section{Introduction}
Recently \cite{lake}, a detailed study of the McVittie solution \cite{mcvittie} was carried out for a non-negative cosmological constant ($\Lambda \geq 0$). The McVittie solution has been known for many years, but it continues to attract interest \cite{newmcvittie}. Even though it is now widely believed that the universe is dominated by a positive cosmological constant, the remarkable  AdS/CFT correspondence \cite{Maldacena} presents a strong argument that the case $\Lambda< 0$ should also be examined. Following \cite{lake} we systematically construct a global view  of the McVittie solution with $\Lambda< 0$ based on numerical integrations of the null geodesics. What results is a situation very distinct from the case $\Lambda \geq 0$: the global structure is completely dominated by a collapse to a Big Crunch, just as in the pure Robertson - Walker case.
\section{The Solution}
\subsection{Overview}
For a perfect fluid with energy density $\rho$ and isotropic pressure $p$ the strong energy condition \cite{hawking} is given by
\begin{equation}\label{tl}
   \rho+3 p \geq 0.
\end{equation}
For a Robertson - Walker background with scale factor $a(t)$, for (\ref{tl}), Einstein's equations with $\Lambda <0$ give
\begin{equation}\label{ray}
    -3\frac{\ddot{a}}{a} \geq -\Lambda > 0
\end{equation}
where $^{.}\equiv d/dt$, and so  we necessarily have a Big Crunch \cite{tipler}. The particular solution we are concerned with here is
the simplest of the McVittie class, and this can be written
in the form (e.g. \cite{nolan}) \cite{notation}
\begin{equation}\label{metric1}
    ds^2=-\left(\frac{1-m/2u}{1+m/2u}\right)^2dt^2+a^2(1+m/2u)^4(dr^2+r^2d\Omega^2_{2})
\end{equation}
where $u\equiv r a$, $m$  is a positive constant and $d\Omega^2_{2}$ is
the metric of a unit 2-sphere. If the McVittie solution (\ref{metric1}) asymptotes to a reasonable Robertson - Walker background, then, as in the pure Robertson - Walker case, $\Lambda < 0$ gives rise to a Big Crunch which dominates the global structure. As previously \cite{lake}, we use the coordinate transformation
\begin{equation}\label{Rdef}
    R(t,r) \equiv u(1+m/2u)^2
\end{equation}
to obtain
\begin{equation}\label{metric2}
    ds^2 = -f(t,R)dt^2-\frac{2H(t)R}{\sqrt{1-2m/R}}dtdR+\frac{dR^2}{1-2m/R}+R^2d\Omega^2_{2}
\end{equation}
where
\begin{equation}\label{fdef}
    f \equiv 1-2m/R-H^2R^2
\end{equation}
and $H$ is the Hubble function $\dot{a}/a$. From (\ref{metric2}) it follows that tangents to surfaces of constant finite
$t$ are spacelike for $R > 2m$ (and so for finite $t$ we set the future orientation $dt/d\lambda>0$ for affine $\lambda$ increasing to the future) and tangents to surfaces of constant $R$ are
spacelike for $f < 0$, null for $f = 0$ and timelike for $f > 0$. We note again that the effective gravitational mass \cite{mass} associated with (\ref{metric2}) is not $m$, but rather $M$, given by
\begin{equation}\label{mass}
    M(t,R)=m+\frac{1}{2}H^2R^3.
\end{equation}
\subsection{The function $H$}
Quite unlike \cite{lake}, we note that because of (\ref{ray}),
\begin{equation}\label{limit}
    t \rightarrow \infty \;\;\; \nexists.
\end{equation}
Rather, we are now interested in models for which
\begin{equation}\label{tt}
    a(0) = a(t_{f})=0, \;\;\; \dot{a}(t_{0})=0
\end{equation}
where
\begin{equation}\label{ttt}
    0\;\;\;<\;\;t_{0}\;\;<\;\;t_{f},
\end{equation}
and
\begin{equation}\label{H}
    H(0<t<t_{0})>0,\;H(t_{0})=0,\;\;H(t_{0}<t<t_{f})<0.
\end{equation}
Note that from (\ref{mass})
\begin{equation}\label{mass0}
    M(t_{0},R)=m,
\end{equation}
and from (\ref{ray})
\begin{equation}\label{hdotneg}
    \dot{H} < 0.
\end{equation}

Whereas from the definition of $u$, $\lim_{t \rightarrow 0, t_{f}}u=0$ for all finite $r$, from the transformation (\ref{Rdef})
\begin{equation}\label{Rlim}
    \lim_{t \rightarrow 0, t_{f}}R=
\begin{cases}
 0 & \text{if }m =0 \\
 \infty & \text{if }m \neq 0\\
\end{cases}\;\;\;.
\end{equation}
As a result, neither $t=0$ nor $t=t_{f}$ are part of the spacetime (\ref{metric2}) for $m \neq 0$. We note that (\ref{Rlim}) shows us that there is no continuous transition from $m \neq 0$ to $m = 0$.
\subsection{Scalar Singularities}
As explained previously \cite{lake}, singularities, as
revealed by scalars polynomial in the Riemann tensor, are reflected here by the Ricci scalar $\mathcal{R}$,
\begin{equation}\label{ricci}
    \mathcal{R}=12H^2+\frac{6\dot{H}}{\sqrt{1-2m/R}},
\end{equation}
since all other invariants, derived
from (partial) derivatives of the metric tensor no higher
than 2, add no new information in the cases under consideration. For $0 < t < t_{f}$, since $\dot{H} \neq 0$, there
are singularities at $R=2m \;(u=m/2)$, which are spacelike. The apparent
singularities at $t = 0$ and at $t=t_{f}$\footnote{We assume that $\dot{a} \neq 0$ at $t=0$ and $t=t_{f}$.}, over the range $2m < R < \infty$ are, as
explained above, not part of the spacetime.
\subsection{The locus $f=0$}
As in the previous analysis \cite{lake}, the locus $f=0$ is important for an understanding of the spacetime (\ref{metric2}). However, due to the nature of the function $H$ studied here, this locus is quite distinct from the locus studied in \cite{lake}. In particular, the roots $R_{\pm}$ studied there do not exist for $\Lambda<0$. First let us note that the locus $f=0$ includes $R=2m$ at $t=t_{0}$ where $H=0$. Moreover, since
\begin{equation}\label{tangent}
    \left(\frac{m}{R^2}-H^2R\right)\dot{R}=H\dot{H}R^2
\end{equation}
along the locus, $R$ can have a vertical tangent in the $R-t$ plane on $0<t<t_{0}$ and $t_{0}<t<t_{f}$ only at $R=3m$. As $R \rightarrow \infty$ the locus becomes $H^2R^2=1$ which requires $H \rightarrow 0$, that is, $t \rightarrow t_{0}^{\pm}$.
\subsection{Null Geodesics - Qualitative}
The radial null geodesics of (\ref{metric2}) satisfy
\begin{equation}\label{geodesic}
    \frac{dR}{dt}=\sqrt{1-2m/R}\left(HR \pm \sqrt{1-2m/R}\right).
\end{equation}
We label the branch ``+" ``outgoing" and the branch
``-" ``ingoing". Clearly
\begin{equation}\label{geodesict0}
    \frac{dR}{dt}\bigg|_{t_{0}}=\pm \left(1-\frac{2m}{R}\right)
\end{equation}
and so the ingoing geodesics already have $dR/dt<0$ at $t_{0}$ whereas the outgoing geodesics have $dR/dt>0$ at $t_{0}$. Further,
\begin{equation}\label{geodesict0f}
    \frac{dR}{dt}=0
\end{equation}
at $f=0$ for ingoing null geodesics when $t < t_{0}$ and for outgoing null geodesics when $t > t_{0}$.
Since $dR/dt>0$ along both branches for $f<0$ and $t<t_{0}$ and $dR/dt<0$ along both branches for $f<0$ and $t>t_{0}$ it follows that the ingoing geodesics reach a maximum $R$ at $f=0$ for $t<t_{0}$ and the outgoing geodesics reach a maximum $R$ at $f=0$ for $t>t_{0}$. A special case is shown below in Figure \ref{figure2} and Figure \ref{figure3}.
\subsection{Energy conditions in general}
Let us start by rewriting the energy density and isotropic pressure in terms of $H$. From Einstein's equations with $\Lambda < 0$ we find
\begin{equation}\label{rhoph}
    8 \pi \rho = 3H^2-\Lambda,\;8 \pi p = -3H^2-\frac{2\dot{H}}{\sqrt{1-2m/R}}+\Lambda.
\end{equation}
As a result, with (\ref{hdotneg}), a general feature of these models is $\rho \geq 0$ and $\rho + p \geq 0$ and so the null and weak energy conditions
are always satisfied. The strong energy condition requires $\rho+3p \geq 0$ which, from (\ref{rhoph}), gives
\begin{equation}\label{strong}
    \frac{\dot{H}}{\sqrt{1-2m/R}}+H^2 \leq \frac{\Lambda}{3} < 0.
\end{equation}
The dominant energy condition requires
\begin{equation}\label{domgen}
    -\rho \leq p \leq \rho.
\end{equation}
It follows from  (\ref{hdotneg}), (\ref{rhoph}) and (\ref{domgen}) that whereas the left-hand inequality is generally satisfied, the right-hand inequality requires
\begin{equation}\label{right}
    -\frac{\dot{H}}{\sqrt{1-2m/R}} \leq  3H^2 - \Lambda.
\end{equation}

\section{A specific form for $H$}
We cannot proceed with further details without a specific form for $H$. For notational convenience, and for a comparison with \cite{lake}, define
\begin{equation}\label{h0}
    H_{0}^{2} \equiv -\frac{\Lambda}{3},\;\; T \equiv 3H_{0}t,
\end{equation}
and take $H_{0}>0$. For $a$ we take the Robertson - Walker scale factor for spatially flat dust with $\Lambda<0$. This gives
\begin{equation}\label{special}
    H=\frac{H_{0}\sin(T)}{1-\cos(T)}.
\end{equation}
Clearly $H$ is periodic with period $2\pi$ and so $T_{0}=\pi$. Moreover,
\begin{equation}\label{hdot}
    \dot{H} = -\frac{3H_{0}^{2}}{1-\cos(T)}.
\end{equation}
The dominant feature in our study is the development of a Big Crunch.
We know that this will occur with $\Lambda <0$ as long as the strong
energy condition holds. For more general equations of state, say $p=\kappa \rho$,
the strong energy condition gives $\rho(1+3 \kappa) \geq 0$. Since the minimum $8 \pi \rho$
in our considerations is $-\Lambda >0$, we would recover the same structure
for $\kappa >-1/3$.  In this sense our choice of $\kappa=0$ is not critical.

\section{Energy conditions}
From (\ref{strong}) we find that the strong energy condition is satisfied for
\begin{equation}\label{strongspecial}
    \frac{3}{\sqrt{1-2m/R}} \geq 2.
\end{equation}
Since the left side of (\ref{strongspecial}) is at least $3$, we see that the strong energy condition is always satisfied. For the dominant energy condition we rearrange (\ref{right}) to give

\begin{equation}\label{diminantsatisfied}
    \frac{R}{m} \geq \frac{8}{3}
\end{equation}
 and so the dominant energy condition fails sufficiently close to the singularities.
\section{Integration of the null geodesics}
\subsection {Integration in the $R - T$ plane}
\begin{figure}[ht]
\epsfig{file=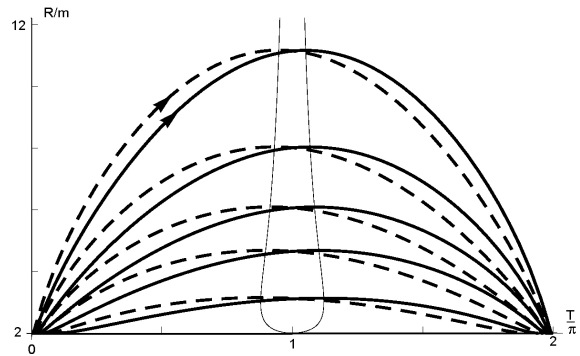,height=2in,width=3in,angle=0}
\caption{\label{figure2}Numerical integrations of (\ref{dimgeo}) and the locus $f=0$. The ingoing geodesics ($``-"$) are shown dashed and reach their maximal value of $R/m$ on the left branch of the locus $f=0$. The outgoing geodesics ($``+"$) are shown solid and reach their maximal value of $R/m$ on the right branch of the locus $f=0$. The future orientation is $T$ increasing as shown. All geodesics begin and end at $R=2m$ which has two distinct parts, separated by the exceptional point $T=\pi, R=2m$. This exceptional point is not part of the spacetime. An enlarged view near $R=2m$ is shown in Figure \ref{figure3}. }
\end{figure}

\begin{figure}[ht]
\epsfig{file=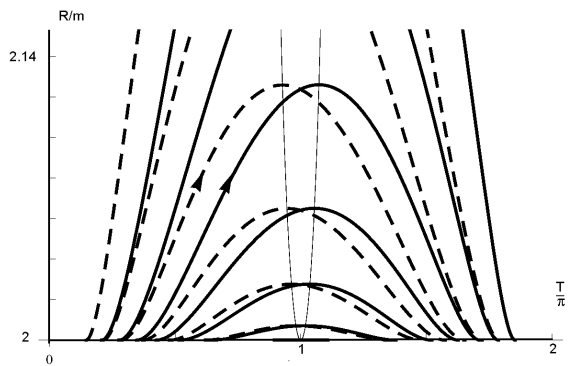,height=2in,width=3in,angle=0}
\caption{\label{figure3} As in Figure \ref{figure2} in the neighborhood of $R=2m$.}
\end{figure}

With (\ref{special}) we can write the null geodesic equations (\ref{geodesic}) in the dimensionless form
\begin{equation}\label{dimgeo}
    \frac{dY}{dT}=\sqrt{1-2/Y}\left(\frac{\sin(T)}{1-\cos(T)}\left(\frac{Y}{3}\right) \pm \frac{1}{3 \delta}\sqrt{1-2/Y}\right)
\end{equation}
where $Y \equiv R/m$ and $\delta$ is the parameter $H_{0} m$. Numerical integrations of (\ref{dimgeo}) are shown in Figures \ref{figure2} and \ref{figure3}.

\subsection {Integration in the $z - T$ plane}
\begin{figure}[ht]
\epsfig{file=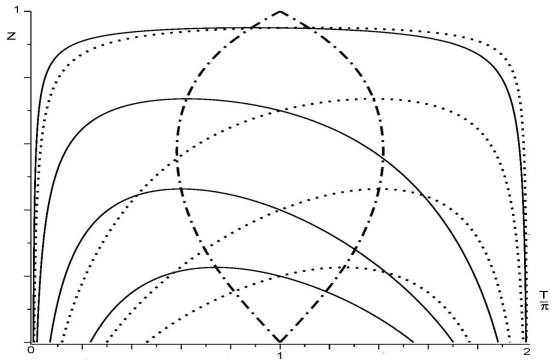,height=2in,width=3in,angle=0}
\caption{\label{figure4}Numerical integrations of (\ref{zform}) and the locus $f=0$. The ingoing geodesics ($``-"$) are now shown solid and reach their maximal value of $z$ on the left branch of the locus $f=0$. The outgoing geodesics ($``+"$) are now shown dashed and reach their maximal value of $z$ on the right branch of the locus $f=0$. The future orientation is $T$ increasing as above. All geodesics begin at $z=0$ for some value of $0 < T < \pi$ and terminate again at $z=0$ for $\pi < T < 2 \pi$. }
\end{figure}

As previously \cite{lake}, we find it numerically convenient to compactify $R$ and define
\begin{equation}\label{z}
    z \equiv \sqrt{1-\frac{2}{Y}}
\end{equation}
so that equation (\ref{dimgeo}) takes the form
\begin{equation}\label{zform}
    \frac{d z}{d T} = \frac{(1 - z^2)}{6} \left(\frac{\sin(T)}{1-\cos(T)} \pm \frac{z(1-z^2)}{2 \delta}\right).
\end{equation}
Numerical integrations of (\ref{zform}) are shown in Figure \ref{figure4}. Any point in the spacetime $(0 < T < 2 \pi, R > 2 m)$ can be connected to the past boundary $(R=2 m, 0< T < \pi)$ by a unique null geodesic from each branch. We use this to construct the conformal diagram below.

\section{Global structure of the spacetime}
\subsection{Construction of the conformal diagram\footnote{Conformal diagrams in the Robertson - Walker case can be found in \cite{gp}.}$^,$\footnote{Our understanding is that null affine distance is relevant iff the
coordinates diverge, indicating the incompleteness of the coordinates for finite affine distances.
There is no divergence of the coordinates in this paper. In particular, if
the boundaries ($R=2m$) are at finite null affine distance, this finiteness
is irrelevant due to the fact that $R=2m$ is genuinely singular and no
extension is possible. If the boundaries ($R=2m$) are at infinite null
affine distance the conformal diagrams remain unchanged and complete. That
is, the affine completeness of the null geodesics is, in this case,
irrelevant, quite unlike the cases $\Lambda \geq 0$ \cite{lake}.}}
We represent the past boundary as a horizontal line in a Cartesian plane ($y=0,-1\leq x\leq 1$), setting the right end of $R=2m$ at $T=0$, and the left end at $T=\pi$. To represent the interval $0<T<\pi$, from $x=-1$ to $1$, we only require a one-to-one function between these two variables. Our choice for $x(T)$ is given in the Appendix. This function was chosen purely for visual reasons by finding the spline curve fit of several points that were adjusted manually to make the resulting conformal diagram more visually appealing. Note that the choice of the shape of the curve ($R=2m, 0<T<\pi$) to be a horizontal straight line, as well as the function $x(T)$ used is arbitrary and does not change the overall global structure presented in the conformal diagram. It is only required that the curve ($R=2m, 0<T<\pi$) be space-like, and the function $x(T)$ be one-to-one. After this point, the procedure we used is identical to that in \cite{lake}.
\subsection{Null geodesics}
\begin{figure}[ht]
\epsfig{file=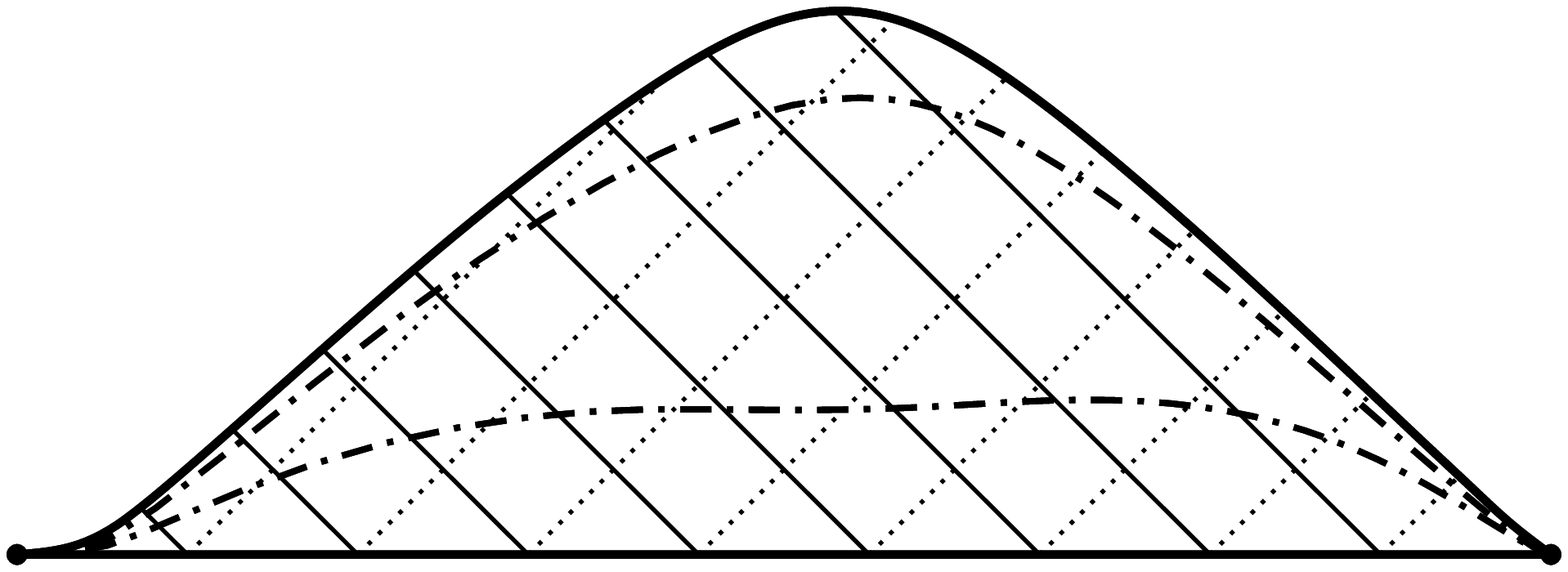,height=2in,width=3in,angle=0}
\caption{\label{nullgeo} Conformal representation of the outgoing null geodesics (dashed) and ingoing null geodesics (solid). The locus $f=0$ is also shown. This locus is globally spacelike. Note that $f>0$ between the two branches of the locus. The boundaries are at $R=2 m$. The point at the lower right is $R \rightarrow \infty, 0 \leq T \leq 2 \pi$. The point at the lower left is $T=\pi, R=2m$.}
\end{figure}

Under the procedure described above, Figure \ref{figure4} is mapped into Figure \ref{nullgeo}.
\subsection{Surfaces of constant $R$ and $T$}
\begin{figure}[ht]
\epsfig{file=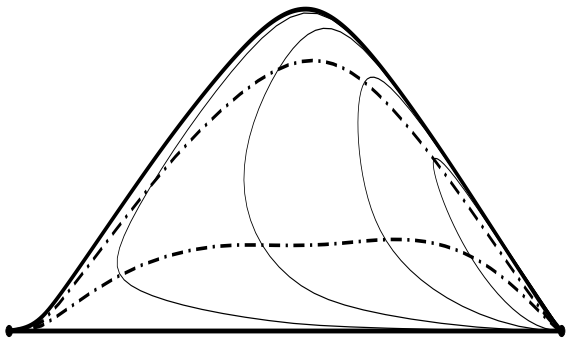,height=2in,width=3in,angle=0}
\caption{\label{consR} The trajectories show surfaces of constant $R$. These are timelike within the locus $f=0$ and spacelike outside the locus. The values of $z$ used to generate these curves are $0.15, 0.3, 0.5$ and $0.7$.}
\end{figure}

\begin{figure}[ht]
\epsfig{file=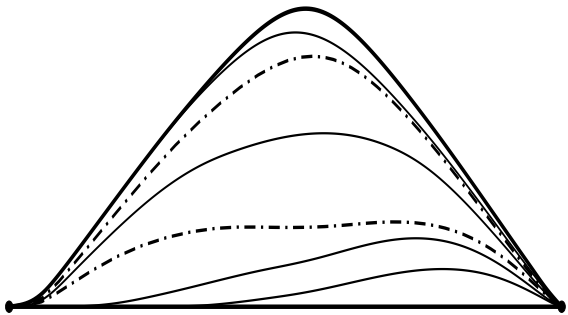,height=2in,width=3in,angle=0}
\caption{\label{consT} The trajectories show surfaces of constant $T$. These are globally spacelike. The values of $T$ used to generate these curves (bottom up) are $\pi/3, \pi/2, \pi$ and $3 \pi/2$.}
\end{figure}

Conformal representations of surface of constant $R$ and constant $T$ are shown in Figures \ref{consR} and \ref{consT} respectively.
\subsection{The fluid streamlines}
\begin{figure}[ht]
\epsfig{file=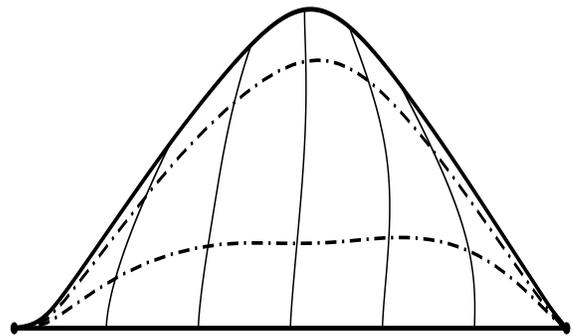,height=2in,width=3in,angle=0}
\caption{\label{fluid}Trajectories of constant $r$ characterized by the constant $\epsilon$ as explained in the text. We note that $\epsilon \rightarrow 1/2 \sqrt[3]{2}$ to the left and $\epsilon \rightarrow \infty$ to the right. }
\end{figure}

The conformal representation of the fluid streamlines $r=constant$ is shown in Figure \ref{fluid}. These trajectories are, of course, globally timelike. From (\ref{Rdef}) we have $u > m/2$ for $R > 2 m$. The scale factor is $a(t)=\mathcal{C}(1-\cos(T))^{1/3}$ where $\mathcal{C}$ is a constant $>0$. Writing $\epsilon =r \; \mathcal{C}/m$ we have
\begin{equation}\label{R/meqn}
    \frac{R}{m} = \frac{(2 \epsilon (1-\cos(T))^{1/3}+1)^2}{4 \epsilon (1-\cos(T))^{1/3}},
\end{equation}
and we note that $R = 2 m$ for $T=T_{0}$ and $T=2 \pi - T_{0}$ where
\begin{equation}\label{Tnot}
    T_{0}=\arccos \left(\frac{8 \epsilon^3-1}{8 \epsilon^3}\right).
\end{equation}
Note that $1/2 \sqrt[3]{2} < \epsilon< \infty$ as explained in the Figure.

\section{Discussion}
Motivated by the AdS/CFT correspondence, we have examined the McVittie solution with a negative cosmological constant $\Lambda < 0$. A detailed construction of the global structure has been given for the case of a background of dust. We have found that the situation is very distinct from the cases $\Lambda \geq 0$ \cite{lake}. As in the pure Robertson - Walker case, we find that $\Lambda<0$ ensures collapse to a Big Crunch, a feature which completely dominates the global structure.

\begin{acknowledgments}
The conscientious efforts of the referee helped us improve the content of this paper. This work was supported in part by a grant (to KL) from the Natural Sciences and Engineering Research Council of Canada. Portions of this work were
made possible by use of \textit{GRTensorII} \cite{grt}.
\end{acknowledgments}

\newpage
\begin{widetext}
\appendix*
\section{$x(T)$}

\begin{equation}
x(T) = 	\left\{\begin{array}{llclll}
       1-2.66T+7.82T^3										& , & 0 		&<T\leq &0.177 \\
       1.02-3.6T+3.43T^2-1.9T^3						& , & 0.177	&<T\leq &0.6 \\
       0.557-0.772T-0.395T^2+0.222T^3			& , & 0.6 	&<T\leq &1.42 \\
       1.87-3.55T+1.56T^2-0.235T^3				& , & 1.42 	&<T\leq &2.2 \\
       -0.601-0.172T+0.0217T^2-0.00230T^3	& , & 2.2 	&<T\leq &\pi
		\end{array} \right.
\end{equation}

\end{widetext}


\begin{thebibliography}{}
\bibitem{notation} We use geometrical units and a signature of +2 throughout.
Functional dependence is designated only on the first
appearance of a function.
\end{thebibliography}
\end{document}